\documentclass[twocolumn]{aastex701}
\usepackage{amsmath}
\usepackage{bm}
\usepackage{subcaption}

\newcommand{\gadget}{\textsc{gadget-4}}

\newcommand{\dd}[1]{\ensuremath{\mathrm{d}#1}}                          
\newcommand{\dnv}[3]{\ensuremath{\frac{\mathrm{d}^#1#2}{\dd{#3}^#1}}}   
\newcommand{\vb}[1]{\ensuremath{\bm{#1}}}                               

\newcommand{\Msun}{\ensuremath{\mathrm{M}_{\sun}}}              
\newcommand{\Gyr}{\ensuremath{\mathrm{Gyr}}}                    
\newcommand{\kpc}{\ensuremath{\mathrm{kpc}}}                    
\newcommand{\rhoSnfw}{\ensuremath{\rho_\mathrm{S, NFW}}}        
\newcommand{\rSnfw}{\ensuremath{r_\mathrm{S, NFW}}}             
\newcommand{\rSp}{\ensuremath{r_\mathrm{S, P}}}                 
\newcommand{\Mvir}{\ensuremath{M_\mathrm{vir}}}                 
\newcommand{\Rvir}{\ensuremath{r_\mathrm{vir}}}                 
\newcommand{\Mext}{\ensuremath{M_\mathrm{ext}}}                 
\newcommand{\Mic}{\ensuremath{M_\mathrm{IC}}}                   

\graphicspath{{./}{figures/}}

\usepackage{xcolor}

\begin{document}

\title{Stellar Cores Live Long and Prosper in Cuspy Dark Matter Halos}

\author[orcid=0009-0001-3652-6878,gname='Jenni',sname='H\"akkinen']{Jenni H\"akkinen}
\affiliation{Department of Physics, University of Helsinki, Gustaf Hällströmin katu 2, FI-00014 Helsinki, Finland}
\email[show]{jenni.hakkinen@helsinki.fi}
\correspondingauthor{Jenni Häkkinen}

\author[orcid=0000-0003-1807-6321,gname='Alexander',sname='Rawlings']{Alexander Rawlings}
\affiliation{Department of Physics, University of Helsinki, Gustaf Hällströmin katu 2, FI-00014 Helsinki, Finland}
\affiliation{Max-Planck-Institut f\"ur Astrophysik, Karl-Schwarzchild-Strasse 1, D-85748 Garching, Germany}
\email{alexander.rawlings@helsinki.fi}

\author[orcid=0000-0003-2403-5358,gname='Till',sname='Sawala']{Till Sawala}
\affiliation{Department of Physics, University of Helsinki, Gustaf Hällströmin katu 2, FI-00014 Helsinki, Finland}
\email{till.sawala@helsinki.fi}

\author[orcid=0000-0003-2496-1925,gname='Matthew',sname='Walker']{Matthew G. Walker}
\affiliation{McWilliams Center for Cosmology and Astrophysics, Department of Physics, Carnegie Mellon University, Pittsburgh, PA 15213, USA}
\email{mgwalker@andrew.cmu.edu}

\begin{abstract}
The existence of cuspy or cored centers of dark matter halos is a crucial discriminant between different dark matter models. It has recently been claimed based on dynamical arguments that perfectly cored stellar systems cannot survive inside cuspy dark matter halos, which would make the observation of stellar cores in ultrafaint dwarf galaxies, where dark matter cores cannot form through baryonic processes, a direct falsification of the cold dark matter paradigm. Here, we use idealized simulations to show explicitly that cored stellar systems like those observed in dwarf galaxies can be stable within cuspy dark matter halos over at least several Hubble times. We also demonstrate that observations of ultrafaint dwarf galaxies cannot distinguish mildly positive, flat, or negative inner density slopes, further precluding the dynamical inference of the gravitational potential from the stellar configuration.
\end{abstract}



\section{Introduction}
Dwarf galaxies, and in particular, the dwarf-spheroidal (dSph) and ultrafaint dwarf (UFD) galaxies surrounding the Milky Way, are among the Universe's most dark matter (DM)-dominated objects. Both in their abundance and in their internal structure, they are key to understanding the nature of DM.

In the standard, cold dark matter (CDM) model,  collisionless DM forms so-called ``cuspy" halos, characterized by a divergent inner density slope that is well described by a Navarro--Frenk--White (NFW) density profile \citep{Navarro-1997}. 
Observational tests of this prediction on scales of individual galaxies have yielded mixed results.  While rotation curves derived from disk galaxies tend to favor ``cored" halos of near-uniform density in their central regions \citep[e.g.,][]{oh11,li20}, the diversity of observed rotation curves includes individual cases that favor NFW-like cusps \citep[e.g.,][]{oman15}.

At the smallest galactic scales, observational tests are further complicated by the fact that the Milky Way's dSph and UFD satellites are pressure-supported systems, precluding the construction and analysis of rotation curves.  Instead, dynamical inferences of their mass distributions are based on statistical analysis of how member stars populate a 6D phase space that is typically observed in only three dimensions (projected position and line-of-sight velocity).

Various techniques have been used to infer the mass distributions within ``classical" dSphs (stellar masses in the range $\sim 10^4\mbox{--}10^7\,\Msun$) from photometric, spectroscopic, and in some cases proper motion data---again with mixed results.  Several studies conclude that the most luminous dSphs (e.g., Fornax and Sculptor) harbor large DM cores, based on Jeans models \citep{Read_2019,Hayashi_2020}, orbit-based Schwarzschild models \citep{Jardel_2012}, direct modeling of the phase-space distribution function \citep[DF;][]{Pascale_2018}, and statistical analysis of chemodynamically distinct subpopulations \citep{Battaglia_2008,Walker_2011,Amorisco_2011,Agnello_2014}. Others, analyzing the same spectroscopic data sets, claim insufficient constraining power to distinguish DM cores from CDM cusps \citep[e.g.,][]{Strigari_2017,Strigari_2018,Zhu_2016}. Including Hubble Space Telescope proper motion data that resolve the internal velocity dispersions of the Draco dSph, \citet{Vitral_2025} constrain any DM core to have radius $\lesssim 1\,\kpc$. Several studies applying homogeneous analyses to all available dSphs conclude that the most luminous ones tend to favor cores, while the least luminous ones favor cusps \citep[e.g.,][]{Read_2019,Hayashi_2020,Hayashi_2023}.  

This scenario is consistent with results of some cosmological and hydrodynamical simulations, which show that supernova feedback from sufficiently vigorous star formation can alter the structure of CDM halos, inducing the formation of cores in what would originally have been cuspy halos \citep{Navarro-1996,mashchenko2008,Pontzen-2012} hosting the most luminous dSphs. This process of astrophysical core formation within CDM halos is expected to become inefficient on scales of the smallest and least luminous UFD galaxies, where feedback from the formation of their paltry stellar populations would have little effect on the structure of the host DM halo \citep[e.g.,][]{madau2014,dicintio2014,onorbe2015,Orkney-2021}.

Thus, the detection of cores in UFDs would directly rule out collisionless CDM. However, the fact that UFDs contain so few stars, usually in a single population, makes the analysis that is already difficult in their brighter counterparts even more so \citep[see, e.g.,][for a review]{simon2019}.

In a series of papers, \citet{almeida_EIM_2024} proposed a new approach to determine the DM density profile of UFDs, requiring only their (projected) surface brightness profiles and applying the Eddington inversion method \citep{almeida2023} to infer the gravitational potential. They concluded that an isotropic stellar population cannot have a cored density profile within a cuspy DM halo, because it implies that the phase-space DF, derived using the Eddington inversion method, has negative values. In particular, \citet{almeida_stellar_2024} claim that the combined surface density profiles of six UFDs require a cored density profile and thus, rule out a cuspy halo, and \citet{almeida_constraints_2025} explore these in the context of self-interacting DM.

Here, we test this proposition in two parts. First, we explicitly test the stability of cored stellar systems inside cuspy DM halos, using idealized numerical simulations. We find that they can be stable in cuspy halos, provided that the two are initially set up in equilibrium. Second, we test the ability to constrain the stellar density profile of UFDs using a finite number of stars. We find that, in particular, the inner slope is degenerate, even before considering observational limitations.

\section{Numerical methods} \label{sec:methods}
Traditional cosmological simulations lack both the mass and the spatial resolution to simultaneously model the DM halos and resolve the stellar systems of UFDs. Furthermore, they require gravitational softening, which could artificially induce or maintain cores in the stellar components \citep[e.g.,][]{Power-2003}.

In this work, we avoid these numerical limitations by representing the DM by an external potential instead of live particles. This allows us to use solar-mass particles and very small softening lengths ($1\mbox{--}10$\,pc, much smaller than the typical core size) for the stellar system. It also greatly reduces the computational cost, allowing us to study a range of parameters and simulate systems over many Hubble times. The choice of an external potential is justified given that the system is completely DM dominated.

\subsection{Initial conditions} \label{subsec:ICs}
We generate the initial conditions (ICs) used to model the dwarf galaxies using the DF method following \citet{hilz2012}. Our fiducial system consists of a stellar component of mass $M_\star=5\times10^3\,\Msun$ embedded in a DM halo with a virial mass of $\Mvir=5\times10^8\,\Msun$. The DM component is modeled by an NFW density profile:
\begin{equation} \label{eq:nfw}
    \rho_\mathrm{NFW} (r) = \frac{\rhoSnfw}{\left(\frac{r}{\rSnfw}\right) \left[1 + \left(\frac{r}{\rSnfw}\right) \right]^2},
\end{equation}
where the characteristic density $\rhoSnfw$ is given by 
\begin{equation}
    \rhoSnfw = \frac{\Mvir}{4\pi \rSnfw^3} \left[\log(1 + c) - \frac{c}{1 + c}\right]^{-1},
\end{equation}
and the NFW scale radius is given by $\rSnfw \equiv \Rvir / c$ with a concentration of $c=15$, typical for this halo mass \citep[e.g.,][]{Diemer-2015} in the $\Lambda$CDM model. For our fiducial model, $\Rvir\simeq16.8\,\kpc$, which corresponds to $\rSnfw\simeq1.12\,\kpc$.

The stellar component is modeled by a Plummer density profile \citep{binney2008}:
\begin{equation} \label{eq:plummer}
    \rho_\mathrm{P} (r) =  \frac{3 M_\star}{4\pi \rSp^3} \left[1 + \left(\frac{r}{\rSp}\right)^2 \right]^{-5/2},
\end{equation}
where $\rSp$ is the Plummer scale radius. Throughout this work, we show results with $\rSp=200\,\mathrm{pc}$. We also repeated the experiment with $\rSp=50\,\mathrm{pc}$ and found qualitatively identical results.

The DF $f_i$ is computed for both components from their respective density profiles $\rho_i$ using Eddington's formula \citep{binney2008}:
\begin{equation}\label{eq:eddington}
    f_i(\mathcal{E}) = \frac{1}{2\sqrt{2}\pi^2} \int_{\Phi_\mathrm{T}=0}^{\Phi_\mathrm{T}=\mathcal{E}} \dnv{2}{\rho_i}{\Phi_\mathrm{T}} \frac{\dd{\Phi_\mathrm{T}}}{\sqrt{\mathcal{E}-\Phi_\mathrm{T}}}.
\end{equation}
Here $\mathcal{E}$ is the relative energy, and $\Phi_\mathrm{T}$ is the total gravitational potential of the system. 
We Monte Carlo sample the DF of the stellar component with particles of $m_\star=1\,\Msun$.
The radial velocity profiles of the stellar particles are ergodic, and the velocity anisotropy $\beta_\mathrm{ani}$, defined as
\begin{equation}
    \beta_\mathrm{ani} \equiv 1 - \frac{\sigma_\theta^2 + \sigma_\phi^2}{2\sigma_r^2},
\end{equation}
where $\sigma_r$, $\sigma_\theta$, and $\sigma_\phi$ are the radial, polar, and azimuthal velocity dispersions, respectively, is zero throughout. Although \citet{almeida_stellar_2024} argue that stellar cores and cuspy halos are incompatible even when this assumption is relaxed, velocity isotropy is chosen as the most stringent test. For our fiducial system, $\sigma_r=\sigma_\theta=\sigma_\phi\simeq5.8\,\mathrm{km/s}$, comparable to values observed for UFDs \citep[e.g.,][and references therein]{simon2019}.

\subsection{Simulations} \label{subsec:simulations}
We run idealized $N$-body simulations using the \gadget{} code \citep{springel_simulating_2021}, compiled using the fast multipole method with second-order multipoles, hierarchical time integration, and an external gravitational field created by a static NFW potential.
The stellar distribution is set up to be in equilibrium with the NFW halo at $t=0$.
In practice, we add to the equations of motion in \gadget{} an additional acceleration term due to an external NFW profile of mass $\Mext = \Mvir$:
\begin{equation}
    \vb{a}_\mathrm{NFW} = \frac{G\Mext}{r^3} \frac{\frac{r}{r + \rSnfw} - \log\left(1 + \frac{r}{\rSnfw}\right) } {\log(1 + c) - \frac{c}{1+c}}\vb{x},
\end{equation}
where $r=|\vb{x}|$ as usual.

We create and perform 10 different realizations of our fiducial system. The number of realizations is chosen to mimic the number of UFDs used in the analysis of \citet{almeida_stellar_2024}. To estimate the effect of the external potential, we also vary its virial mass, $\Mext$, (and radius) from that of one of the initial fiducial systems, $\Mic$, and perform seven simulations with an unequal mass ratio $\Mext / \Mic$. In other words, in these simulations, the initial stellar system is explicitly out of equilibrium with the potential in which it will be evolved.

We investigate the effect of softening by performing simulations with softening lengths of 1, 10, and 100\,pc. There was no clear distinction between these values, and we chose a softening of 10\,pc to be used for this work. The scale at which the gravitational force is affected by softening is $r_\mathrm{S, G} = 28\,\mathrm{pc}$ \citep{springel_simulating_2021}, much less than the size of the stellar core.

Due to the low number density of stars, both the time between stellar collisions and the relaxation time \citep[e.g.,][]{Errani-2025} are much longer than the Hubble time, justifying the collisionless approximation implicit in our simulations. We also note that the total kinetic energy of the stars is $\sim1000$ times below the binding energy of the cusp. Hence, any possible response of the DM to the stars, which our static potential would fail to capture, is guaranteed to be negligible.

To study both the short-time evolution and the long-term stability, the simulations are analyzed at $t=0\,\Gyr$, at 100 additional outputs with a cadence of $\Delta_t = 10\,\mathrm{Myr}$ until $t=1\,\Gyr$, and at 99 additional outputs with a cadence of $\Delta_t = 1\,\Gyr$ until 100\,\Gyr. The dynamical time of stars at the stellar half-mass radius ($\sim0.27\,\mathrm{kpc}$) in our system is $\sim25\,\mathrm{Myr}$, much less than the total simulation time.

\section{Evolution}\label{sec:evolution}

\begin{figure*}[ht!]
    \centering
    \includegraphics[width=7in]{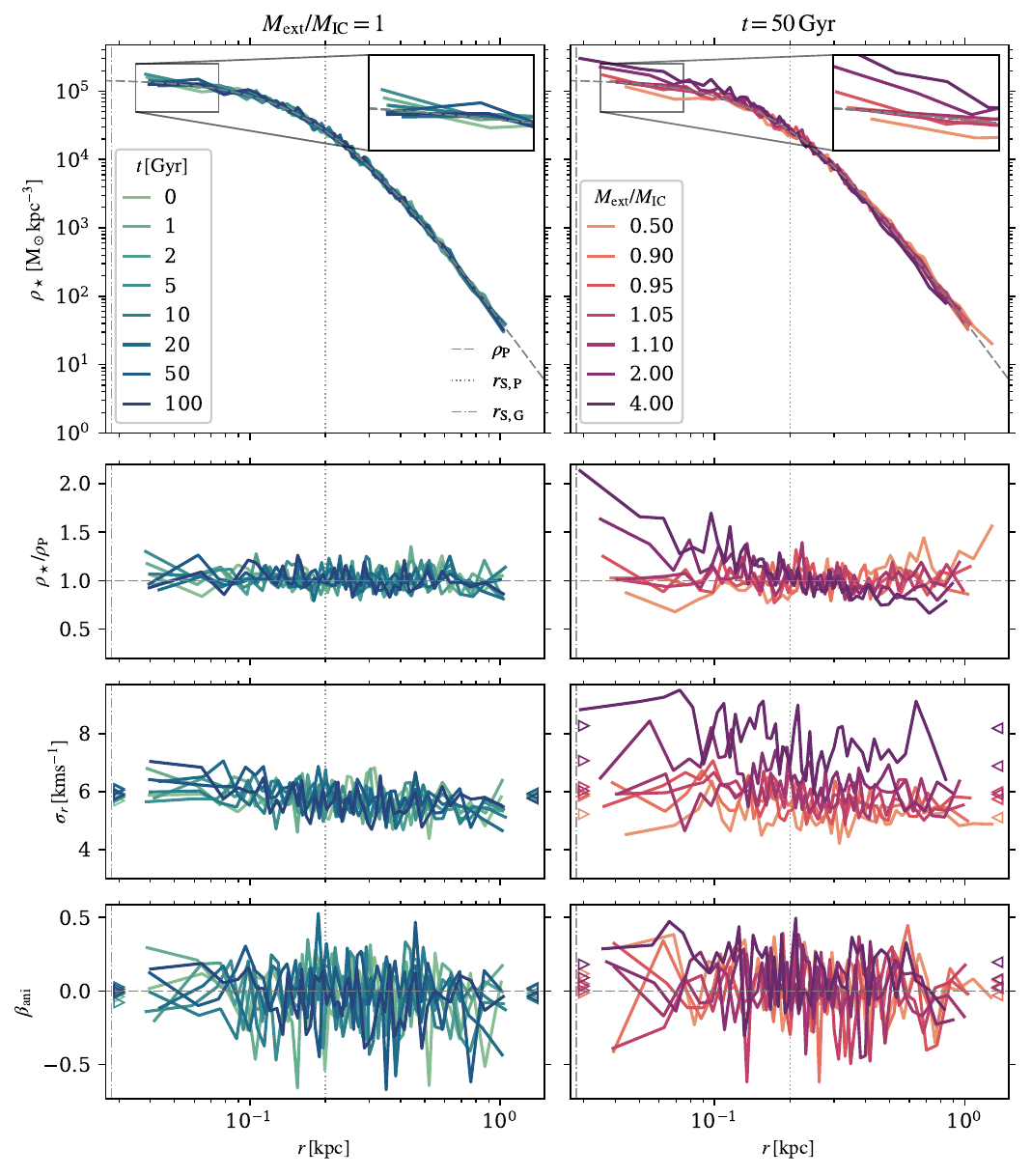}
    \caption{Evolution of the binned stellar density $\rho_\star$, velocity dispersion $\sigma_r$, and velocity anisotropy $\beta_\mathrm{ani}$ as functions of radius $r$. The dotted line shows $\rSp=0.2\,\kpc$, and the dash-dotted line shows $r_\mathrm{S, G}=0.028\,\kpc$. \textit{Left column:} one simulation of the fiducial system for selected times. \textit{Right column:} seven simulations with unequal mass ratios $\Mext/\Mic$ at $t=50\,\Gyr$. \textit{First row:} stellar density profile, with the zoomed-in region showing the inner region within $r=0.075\,\kpc$, and the dashed line showing the analytic Plummer profile $\rho_\mathrm{P}$. \textit{Second row:} the ratio of $\rho_\star$ to $\rho_\mathrm{P}$, with the dashed line showing $\rho_\star / \rho_\mathrm{P} = 1$. \textit{Third row:} velocity dispersion profile, with the triangles showing the time-average of $\sigma_r$ within (on the left) and outside (on the right) $\rSp$. \textit{Fourth row:} velocity anisotropy profile, with the dashed line showing $\beta_\mathrm{ani} = 0$, and the triangles showing the time-averaged $\beta_\mathrm{ani}$.}
    \label{fig:evolution}
\end{figure*}

\begin{figure*}[ht!]
    \centering
    \includegraphics[width=7in]{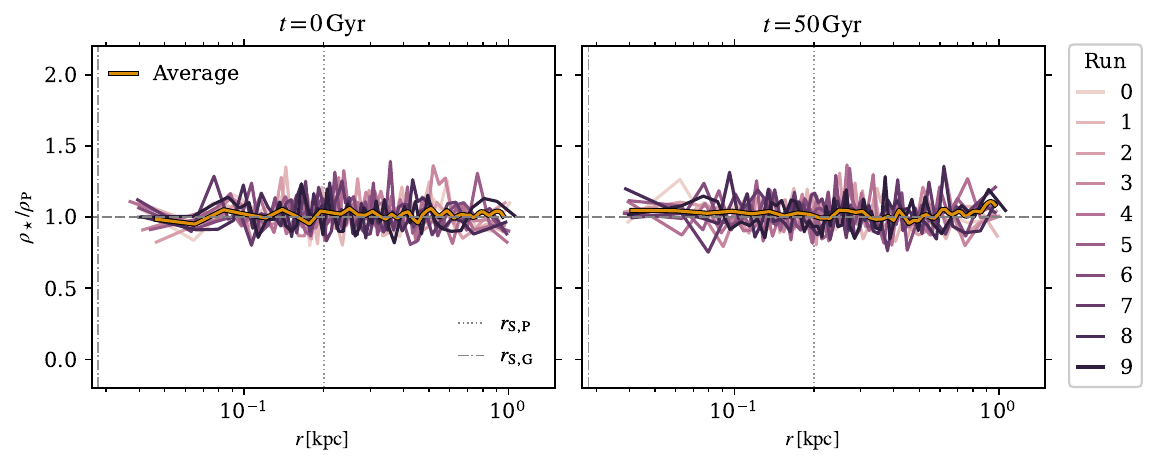}
    \caption{Similar to the left panel of the second row of \autoref{fig:evolution}, but for 10 realizations of the fiducial system, at times $t=0\,\Gyr$ (\textit{left}) and $t=50\,\Gyr$ (\textit{right}). The orange line shows the average ratio of all runs.}
    \label{fig:fiducials}
\end{figure*}

\begin{figure*}[ht!]
    \centering
    \includegraphics[width=7in]{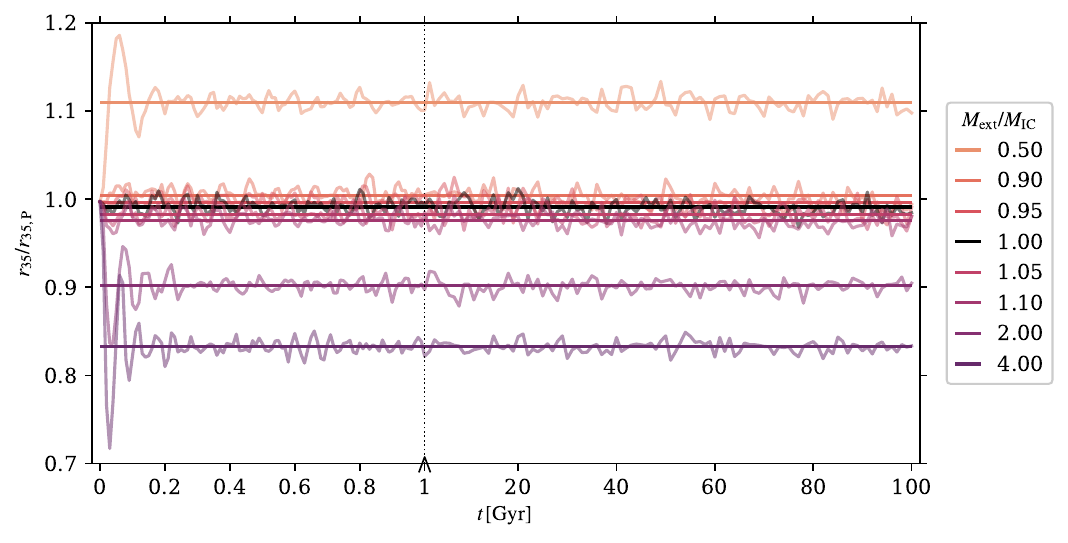}
    \caption{The ratio of the Lagrangian radii $r_{35}$ in simulations to that of the Plummer profile $r_{35, \mathrm{P}}$, as a function of time $t$ for eight different mass ratios $\Mext / \Mic$. The lighter fluctuating lines show the evolution at each output, while the darker horizontal lines corresponds to the time-averaged value across the whole simulation. The $t$-axis is divided into two ranges separated by the black dotted line: the left side shows the evolution within the first 1\,\Gyr\ and the right side within the remaining 99\,\Gyr.
    }
    \label{fig:r35}
\end{figure*}

To calculate the evolution of the stellar density profiles from simulation data, we first center the system using the shrinking-sphere method \citep{Power-2003} implemented in \textsc{pynbody} \citep{pynbody} and remove the outermost 250 (5\%) of stellar particles. In the outskirts of the system, the sampling is poor, and the estimated density can fluctuate significantly, as the random motions of individual particles can cause a net increase of the maximum radii and hence, an underestimate of the density. We note that in observed UFDs, the outermost 1\%--10\% of stars are also typically not detectable \citep[e.g.,][]{richstein2024} due to the low surface brightness and foreground contamination. We bin the remaining particles in 50 radial shells of equal particle number, each containing 95 particles (95\,\Msun). We explicitly test the effect of using different numbers of bins in \autoref{app:bin}. 

\autoref{fig:evolution} shows the evolution of the density, velocity dispersion, and velocity anisotropy profiles of the Plummer stellar system in the external, cuspy gravitational potential. The left column shows the profiles measured from the ICs at $t=0\,\Gyr$ and at seven later times up to $t=100\,\Gyr$, for one fiducial system, evolved in the same external potential that was used to generate the ICs. While the density measured in the simulation fluctuates around the Plummer profile from which the ICs were constructed, as seen in the first two rows of the figure, there is no discernible evolution. In particular, there is no evidence for a change in the inner density profile induced by the evolution in the cuspy gravitational potential.

The right column shows the profiles measured at $t=50\,\Gyr$, in seven simulations which use identical ICs to the one shown on the left but different external potentials. In this case, the density profile can change, with an increased inner stellar density obtained when the external potential is deeper and a decreased one when it is shallower than the potential used to generate the ICs. However, provided that the ICs are created in equilibrium with the external potential, the stellar core is maintained.

The left panel in the third row of \autoref{fig:evolution} shows how the velocity dispersion of the fiducial system fluctuates around the mean value of $\sigma_r\simeq5.8\,\mathrm{km/s}$, with a slight decrease towards the outer region of the system, similar to observations of UFDs. As expected, the right panel shows how deeper or shallower external potentials result in increased or decreased stellar velocity dispersions, respectively.

As noted by \citet{almeida_EIM_2024}, circular orbits $(\beta_\mathrm{ani} \rightarrow - \infty)$ can accommodate any combination of density and potential, including a cored stellar density in an NFW potential, but those are not physically motivated. Our systems are initially set up with $\beta_\mathrm{ani} = 0$, and their evolution from this can be seen in the bottom row of \autoref{fig:evolution}. As we will show in more detail below, changing the external potential ($\Mext/\Mic \neq 1$) induces net radial motions, which can induce (positive) velocity anisotropy. However, provided that they are evolved in the same potential in which they were created ($\Mext/\Mic=1$), our systems maintain isotropy throughout their evolution.

The fluctuations in the density profile shown in \autoref{fig:evolution} arise simply from the fact that the finite number of $N=5000$ stars provide only a sample of the underlying distribution:~observations of UFDs, which often contain only of the order of hundreds of stars \citep[e.g.,][]{simon2019}, are even more susceptible to sampling variance. This effect is demonstrated in \autoref{fig:fiducials}, which shows the stellar density profiles of 10 different realizations of the fiducial system, both in the ICs and at $t=50\,\Gyr$. Each individual profile fluctuates around the Plummer value, with $\rho_\star / \rho_P \sim 10\%$. The average of all 10 realizations is close to the Plummer value. There is no discernible evolution, neither in the average nor in any of the individual profiles.

A more fine-grained view of the evolution is shown in \autoref{fig:r35}, which compares the radius that encloses 35\% of stellar particles, $r_{35}$, in simulations to that of the analytic Plummer profile, $r_{35,\mathrm{P}}$, which is approximately equal to the scale radius of the ICs ($r_\mathrm{S,P}=0.2\,\kpc$). The simulations shown here are the same ones as in \autoref{fig:evolution}.

It can be seen that evolving the stellar distribution in an external potential much shallower than that used to create the ICs leads to a rapid expansion of the core, followed by a damped oscillation, before a new equilibrium is obtained in a more extended stellar system. Conversely, an external potential much deeper than that used to create the ICs leads to a rapid collapse, a damped oscillation, and finally a more compact stellar system. In each case, the new equilibrium is reached in less than 200\,\ensuremath{\mathrm{Myr}}.

Due to the finite sampling, random fluctuations in $r_{35}$ persist throughout the remaining 100\,\Gyr, making the evolution of systems with small changes to the external potential difficult to distinguish. For this reason, we also show the time-averaged values of $r_{35}$, which demonstrate these differences better. The time-averaged value of the fiducial system ($\Mext/\Mic=1$) is very slightly below $r_{35}/r_\mathrm{35, P}=1$, which can be considered to be caused by sample variance.

In summary, while stellar systems do respond to changes in the external potential, stellar cores can be maintained within cuspy DM halos for many Hubble times.

\section{Stellar density inference}\label{sec:fitting}

\begin{figure*}
    \centering
    \includegraphics[width=7in]{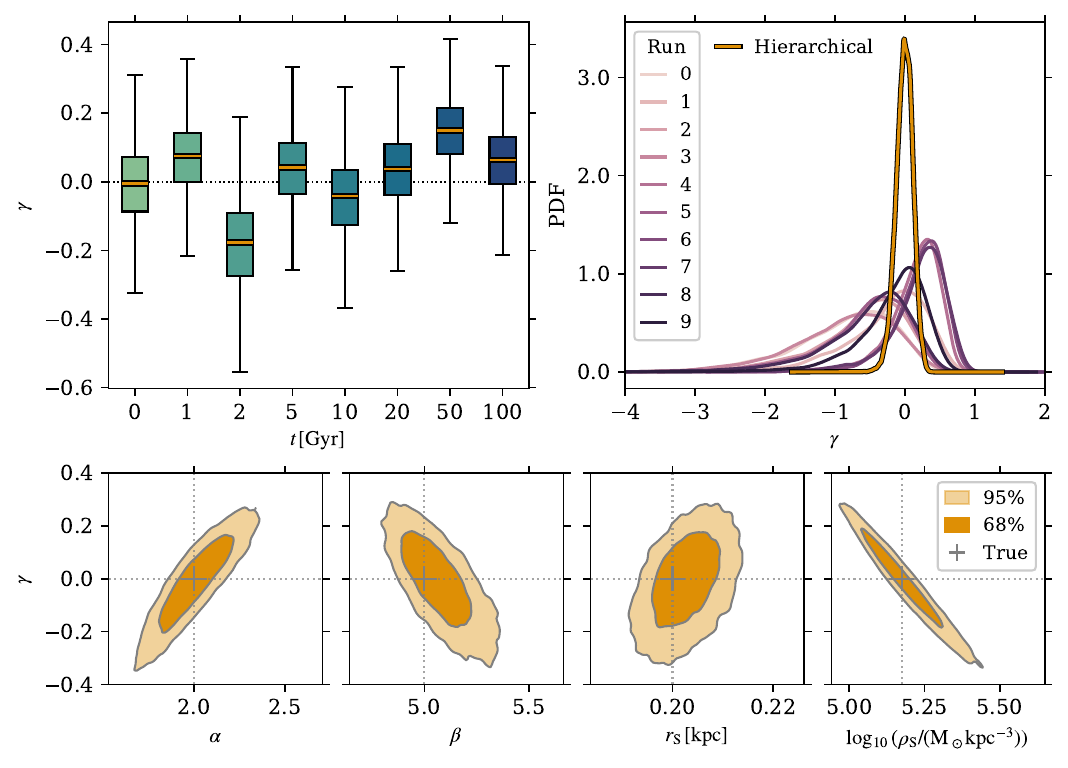}
    \caption{Results of the stellar density fitting. \textit{Top-left panel:} the distribution of $\gamma$ for selected times (as in the left column of \autoref{fig:evolution}) for the hierarchical fit. The orange lines show the median of the fit at each time, and the dotted line shows $\gamma=0$. \textit{Top-right panel:} the PDF of $\gamma$ at $t=0\,\Gyr$. The orange line shows the hierarchical fit, and the remaining lines show the nonhierarchical fits for the 10 fiducial simulations. \textit{Bottom panels:} correlation of $\gamma$ with the parameters $\alpha,\, \beta,\, r_\mathrm{S},\, \mathrm{and}\, \rho_\mathrm{S}$, respectively. The contours show the 68\% and 95\% confidence intervals, and the dashed crossing lines show the true parameters of the underlying Plummer distribution.}
    \label{fig:fitting}
\end{figure*}

Having demonstrated that cored stellar systems can, in fact, be stable and long-lived in cuspy DM halos, we will revisit the previous research which had suggested the opposite conclusion.

We note that a crucial assertion when applying the Eddington inversion method to rule out cuspy DM halos is the presence of stellar cores, and more specifically, the presence of nonnegative inner density slopes. In \citet{almeida_stellar_2024}, this assertion is supported by fitting the combined projected surface density profiles of six UFDs from \citet{richstein2024} and inferring the corresponding phase-space DF. The profiles for all six UFDs are assumed to have the same polynomial shape, which is found to have an inner plateau with a near-zero slope. In an NFW potential, the best-fitting DF producing these profiles is found to contain negative parts, leading the authors to conclude such potential to be inconsistent with the observations. Here, we test the ability to distinguish different inner density slopes with our simulated stellar Plummer profiles in the fiducial setup ($\Mext = \Mic$).

We fit at times $t=(0, 1, 2, 5, 10, 20, 50, 100)\,\Gyr$ the full 3D information of the stellar distribution with the $(\alpha, \beta, \gamma)$ double-power-law profile \citep{merritt2006,dicintio2014}:
\begin{equation}\label{eq:abg}
    \rho(r) = \frac{\rho_\mathrm{S}}{\left( \frac{r}{r_\mathrm{S}} \right)^\gamma \left[1 + \left(\frac{r}{r_\mathrm{S}} \right)^\alpha\right]^{(\beta-\gamma)/\alpha}}.
\end{equation}
Here $\rho_\mathrm{S}$ is a characteristic density, and $r_\mathrm{S}$ is a characteristic scale length. The inner and outer regions have logarithmic slopes of $-\gamma$ and $-\beta$, respectively, with the transition between these regions governed by $\alpha$. A Plummer sphere (\autoref{eq:plummer}) has $(\alpha, \beta, \gamma)=(2,5,0)$, whereas for an NFW profile (\autoref{eq:nfw}), the parameterization is $(1,3,1)$. We fit for all free parameters in \autoref{eq:abg} ($\alpha, \beta, \gamma, r_\mathrm{S}, \mathrm{and}\, \rho_\mathrm{S}$).

To infer the analytic stellar density profile from the simulation data, we fit the double-power-law profile to the stellar density, using the same binning as discussed in \autoref{sec:evolution}.

We use hierarchical Bayesian inference to jointly fit the density profiles of all 10 fiducial runs at the aforementioned times: a detailed methodology is provided in \autoref{app:fits}.
This way, we avoid many of the problems---such as overfitting, loss of data heterogeneity, and diminished precision of parameter estimates for small sample sizes---that are commonly associated with inference performed when completely stacking the data into a single set or performing inference on related datasets independently \citep{gelman2006}.

An important finding of our analysis is the inability to constrain $\gamma$ as strictly positive at any time, indicating that despite \textit{a priori} knowledge that the derivative of the stellar density profile from which our samples are drawn is always $\dd{\rho}/\dd{r} \leq 0$, this cannot be unequivocally said from the measured density profile.
By maximizing the data available in the inference, we constrain the population mean and standard deviation of the inner density slope at $t=0\,\Gyr$ as $\mu_\gamma=-0.005_{-0.208}^{+0.178}$ and $\sigma_\gamma=0.011_{-0.010}^{+0.026}$, respectively, where the uncertainties refer to the 90\% confidence interval of the parameter.
The time evolution of the full distribution of $\gamma$ is shown in the top-left panel of \autoref{fig:fitting}.
At all times, $\gamma$ is consistent with a value of 0; however the spread in the distributions means that $\gamma<0$ could be easily inferred: we find that the probability $P(\gamma\leq0)=0.52$.

For comparison, we fit a nonhierarchical model independently to the 10 fiducial simulations at $t=0\,\Gyr$ (i.e., we fit for the free parameters in the double-power-law profile directly).
Unsurprisingly, the constraints on $\gamma$ are significantly weaker when inferring the density model parameters independently for each simulation compared to the partial stacking allowed for by the hierarchical model, as shown in the top-right panel of \autoref{fig:fitting}. While all individual realizations are drawn from the same underlying Plummer distribution with $\gamma = 0$, due to sample variance, the probability distribution function (PDF) inferred from each realization differs. In particular, in every realization, the PDF extends to both positive and negative values of $\gamma$.

To demonstrate degeneracies between the parameters of the fitted double-power-law profile, in the bottom panels of \autoref{fig:fitting}, we show contours of the 68\% and 95\% confidence intervals of the inferred parameters for the hierarchical model. We see that their true values ($\alpha=2,\,\beta=5,\,\gamma=0,\,r_\mathrm{S}=0.2\,\kpc,\,\mathrm{and}\,\rho_\mathrm{S}\simeq1.5\times10^5\,\Msun\kpc^{-3}$) are recovered very well in this case. However, $\gamma$ is strongly correlated with the other parameters, especially with $\rho_\mathrm{S}$ but also clearly with $\alpha$ and $\beta$. This further shows the difficulty in constraining the inner density slope precisely:~for instance, even a slight increase in $\rho_\mathrm{S}$ favors a more positive inner slope ($\gamma<0$), and vice versa. In other words, the degeneracy of the free parameters in the profile can easily mislead the prediction of the inner density slope.

\section{Discussion}\label{sec:discussion}

When we discuss cores, we use the widely adopted definition of density profiles that are asymptotically flat, i.e., whose inner slope $\dd \rho/\dd r \rightarrow 0$ as $r \rightarrow 0$. That is, we consider a density profile such as the Plummer profile (\autoref{eq:plummer}), or the double-power-law profile (\autoref{eq:abg}) with $\gamma=0$, to be cored, even though $\dd \rho/\dd r < 0$ for any finite $r$. Applying the Eddington inversion method to rule out cuspy potentials may require a more limited definition ($\dd \rho/\dd r > 0$ at some finite radius) by which Plummer profiles would not be considered cored. However, we have shown that for typical UFDs, measurements of the true inner slope from small numbers of stars are ambiguous.

A known constraint of the Eddington inversion method \citep{almeida_EIM_2024} is that perfect stellar cores with $\gamma\le0$ are incompatible with NFW potentials when the velocity distribution is isotropic or radially biased ($\beta_\mathrm{ani}\geq0$). The incompatibility arises from the method producing a DF that is not everywhere nonnegative (i.e., there exists an $\mathcal{E}$ such that $f(\mathcal{E})<0$) and is hence unphysical.

If we consider our ICs, which are by definition ergodic and spherically symmetric, the DF is solely a function of the relative energy (as in \autoref{eq:eddington}).
Effectively, $f$ tells us the probability of finding a stellar particle in the system with an energy in the infinitesimal volume $\dd{\mathcal{E}}$ about $\mathcal{E}$.
In practice, however, we are unable to numerically sample from such an infinitesimal volume, and rather sample from a \textit{coarsened} DF $\bar{f}(\mathcal{E})$ \citep{binney2008} that defines a finite volume of width $\Delta\mathcal{E}$, where $\bar{f}$ is the expectation value, or mean, of $f$ within the volume $\Delta\mathcal{E}$.
Critically, whilst $\bar{f}(\mathcal{E})$ may be everywhere nonnegative in the volume $\Delta\mathcal{E}$, $f(\mathcal{E})$ in a subregion $\dd{\mathcal{E}} \in \Delta\mathcal{E}$ may not be.
The task then becomes ensuring that our approximation
\begin{equation}
    \lim_{N\rightarrow\infty} \sum_i^N \bar{f}(\mathcal{E}_i) \rightarrow \int_\Omega f(\mathcal{E}) \dd{\mathcal{E}}
\end{equation}
is a valid one over the domain $\Omega$, so that the structure of $\bar{f}$ captures the overall structure of $f$.

In creating our ICs, we calculate $f$ at $2\times10^3$ equally-spaced values of $\mathcal{E}$:~this defines our coarsened DF\footnote{Note that we use interpolation so as to evaluate our DF at any $\mathcal{E}$, and not just at the values used to construct the DF.}.
We have tested the effect of increasing the number of $\mathcal{E}$ values $f$ is evaluated at by a factor of 10, and performed a two-sample Kolmogorov--Smirnov (KS) test with a two-tailed null hypothesis to assess if there is a meaningful difference between evaluations of the DF.
We find that the test statistic is $t_\mathrm{KS}=9.5\times10^{-4}$ and the $p$-value is $\sim 1.0$:~the DFs are indistinguishable.
Consequently, we are confident that our constructed $\bar{f}$ is a valid approximation to the true $f$. 
This is further supported by the constructed ICs agreeing well with the desired Plummer profile (shown in \autoref{fig:fitting}), even with no particles exactly in the center requiring high enough energies to result in a negative value being sampled for the DF\footnote{Sampling a particle with such high energy is possible but highly unlikely.}. While negative DFs are clearly unphysical, analytic DFs are only descriptive, which real and finite systems are not required to fully sample.

Although the combination of perfectly cored density components and perfectly cuspy potentials may lead to partially negative DFs, this does not necessarily lead to real-world contradictions. Any smooth density profile (or DF) is only an approximative description of a finite stellar system, rather than its inherent properties. It is worth pointing out that the “cusp” in an NFW-halo is itself unphysical, but nevertheless, the model provides a good description of the measurable DM distribution. The important question is not whether a model is unphysical at its extremes, but whether it successfully describes the physical reality.

\section{Conclusions}\label{sec:conclusions}
The existence of cored rather than cuspy DM profiles in UFDs would have profound implications, effectively falsifying the standard cosmological model. However, our results challenge previous assertions of evidence for DM cores based on cored stellar density profiles. 
First, we have shown that cored stellar systems can persist within cuspy halos, provided that they have formed in initial equilibrium, as would be expected for a stellar system forming within an existing DM halo in the standard galaxy formation paradigm.

Second, we have demonstrated the limitations of determining the crucial inner slope of the stellar density profiles of UFDs. Even with a conservative sample size and in the absence of observational errors, it is exceptionally difficult to distinguish in UFDs a mildly positive, negative, or zero inner density gradient. 
This is heightened by the small collection of stars that constitute such a dwarf galaxy, increasing the observed variance in stellar mass density profiles. While future observations may yet lead to a viable application of the Eddington inversion method, we conclude that based on current observations, the existence of stellar cores in UFDs provides no evidence against cuspy DM halos.

\begin{acknowledgments}
We thank the reviewer for their valuable feedback that helped us improve our manuscript. J.H.~acknowledges the support of the Magnus Ehrnrooth Foundation. A.R.~acknowledges the support of the University of Helsinki Research Foundation. J.H.~and T.S.~acknowledge support by Research Council of Finland grants 354905 and 339127. J.H., A.R., and T.S.~acknowledge support by ERC Consolidator Grant KETJU (no.~818930).  M.G.W.~acknowledges support from the National Science Foundation (NSF) grant AST-2206046. This work used facilities hosted by the CSC -- IT Centre for Science, Finland. Open access funded by Helsinki University Library. \vfill\null
\end{acknowledgments}

\begin{contribution}

\textbf{JH}: data curation, formal analysis, investigation, methodology, visualization, writing – original draft. 
\textbf{AR}: formal analysis, investigation, methodology, software, writing – original draft. 
\textbf{TS}: conceptualization, supervision, writing – original draft. 
\textbf{MGW}: conceptualization, writing – original draft.


\end{contribution}

%

\software{\gadget{} \citep{springel_simulating_2021}, 
          \textsc{pynbody} \citep{pynbody},
          NumPy \citep{harris_array_2020},
          SciPy \citep{virtanen_scipy_2020},
          Matplotlib \citep{hunter_computing_2007},
          \textsc{Stan} \citep*{standevelopmentteam2018},
          CmdStanPy \citep*{standevelopmentteam2018},
          Arviz \citep{kumar2019}.
          }


\appendix
\section{Binning of stellar particles}\label{app:bin}
We test the effect of using different numbers of bins when constructing radial profiles on our conclusions, in particular for the inference of the $(\alpha, \beta, \gamma)$ double-power-law profile.
In all cases, we use bins of equal particle counts (as opposed to equal bin widths) to ensure sufficient and comparable sample sizes within each bin from which we determine statistics, such as the mean density within a bin. 
In the following, we discuss signal-to-noise (S/N) in the Poisson sense, i.e.,~$\mathrm{S/N} = N/\sqrt{N} = \sqrt{N}$.

We first note that determining an optimal binning strategy is a result of two competing factors \citep[e.g.,][]{tarumi2021}: 
increasing S/N, achieved with increasing the number of particles per bin, and increasing spatial resolution, achieved with increasing the total number of bins.
Motivated by observational campaigns of UFDs \citep[e.g.,][]{drlica-wagner2015,tan2025}, we require that the minimum S/N in each radial bin is $\mathrm{S/N} \simeq 10 \implies N\simeq 100$.
We thus try binning the inner 95\% of stellar particles using 5, 10, 25, and 50 bins of equal particle counts\footnote{We also tested a case using 95 bins, however due to the low S/N the profile fits were unable to be recovered.}. 
Already we note that using 5 bins, whilst achieving a high S/N of $\sim30$, barely resolves the inner regions of the stellar profile $r<r_\mathrm{S}$, thus setting a lower limit to the number of bins.
Using 50 bins achieves our desired $\mathrm{S/N}\simeq10$, thus setting our upper limit.

For each binning strategy, we attempt to infer the parameters $\alpha$, $\beta$, $\gamma$, and $r_\mathrm{S}$ of the double-power law model from our ICs (see \autoref{app:fits}), which we know \textit{a priori} to be 2, 5, 0, and 0.2, respectively.
We then determine the Mahalanobis distance between the posterior median of the parameters and the known values: unlike the Euclidean measure, the Mahalanobis measure accounts for the covariance $V$ between the parameters.
The Mahalanobis distance is defined as
\begin{equation}
    d_\mathrm{M} = \sqrt{\left(\theta_\mathrm{posterior} - \theta_\mathrm{true} \right) V^{-1}\left(\theta_\mathrm{posterior} - \theta_\mathrm{true} \right)^\top}.
\end{equation}
A smaller $d_\mathrm{M}$ is indicative of a better recovery of the \textit{a priori} truth than a higher $d_\mathrm{M}$.
We find that as a function of bin count, $d_\mathrm{M}$ decreases exponentially as $d_\mathrm{M} \propto N_\mathrm{bin}^{-0.43}$: from $N_\mathrm{bin}=25$ to $N_\mathrm{bin}=50$, the difference in distance measures is only $\Delta d_\mathrm{M}=0.24$, indicating the two strategies recover the \textit{a priori} truth comparatively well. 
We thus adopt the binning that provides the higher spatial resolution, namely $N_\mathrm{bin}=50$ with 95 particles in each bin, as the chosen binning strategy in this work.

\section{Fitting density profiles}\label{app:fits}
\begin{figure}
    \centering
    \includegraphics[width=0.8\textwidth]{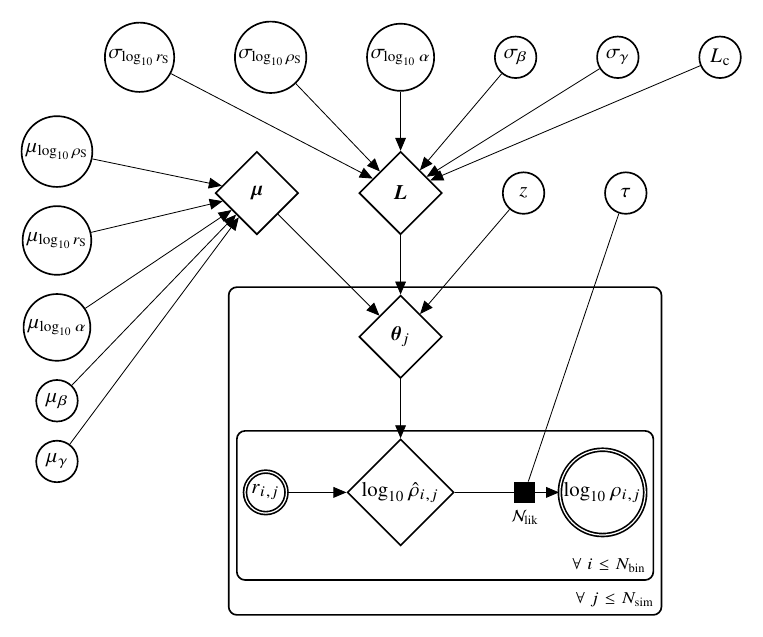}
    \caption{
        Directed acyclic graph showing the model parameters in the fit to the density profile.
        Single-line circle nodes represent fit parameters, double-line circles represent measured quantities (the data), and diamond nodes represent deterministic quantities.
        Parameters outside the inner rectangle do not depend on the measured radius.
    }
    \label{fig:dag}
\end{figure}

\begin{table}
    \caption{Weakly informative prior distributions for the $(\alpha, \beta, \gamma)$ double power-law density profile. 
    The parameter distributions refer to the single-line circle node variables in \autoref{fig:dag}.}
    \label{tab:abg_hyperpars}
    \centering
    \begin{tabular}{llc}
        \hline
        Hyperparameter & Distribution & Truncation \\
        \hline
        $\mu_{\log_{10}\rho_\mathrm{S}}$ & $\mathcal{N}(5, 1)$ & $-5 \leq x \leq 10$ \\
        $\sigma_{\log_{10}\rho_\mathrm{S}}$ & $\mathcal{N}(0, 1)$ & $x > 0$ \\
        $\mu_{\log_{10}r_\mathrm{S}}$ & $\mathcal{N}(0, 1)$ & None \\
        $\sigma_{\log_{10}r_\mathrm{S}}$ & $\mathcal{N}(0, 0.5)$ & $x > 0$ \\
        $\mu_{\log_{10}\alpha}$ & $\mathcal{N}(0, 0.5)$ & None \\
        $\sigma_{\log_{10}\alpha}$ & $\mathcal{N}(0, 0.5)$ & $x>0$ \\
        $\mu_{\beta}$ & $\mathcal{N}(0, 4)$ & None \\
        $\sigma_{\beta}$ & $\mathcal{N}(0, 2)$ & $x>0$ \\
        $\mu_\gamma$ & $\mathcal{N}(0, 2)$ & None \\
        $\sigma_\gamma$ & $\mathcal{N}(0, 2)$ & $x > 0$ \\
        $\tau$ & $\mathcal{N}(0, 1)$ & $x>0$ \\
        $L_\mathrm{c}$ & LKJCholesky$(2)$ & None \\
        $z$ & $\mathcal{N}(0, 1)$ & None \\
        \hline
    \end{tabular}
\end{table}

We fit the $(\alpha, \beta, \gamma)$ double-power-law profile using hierarchical Bayesian inference implemented in \textsc{Stan} \citep*{standevelopmentteam2018}, with four chains of 2000 posterior draws each (i.e., post-burn-in period).
Recognizing that the 10 fiducial simulations describe the same physical system, albeit with different realizations of the particle sampling, we can jointly model the individual density profile parameters, denoted $\bm{\theta} \equiv \left[ \log_{10}\rho_\mathrm{S}, \log_{10}r_\mathrm{S}, \log_{10}\alpha, \beta, \gamma \right]$, as realizations of a \textit{population} of possible density profiles that satisfy the exchangeability criterion \citep{gelman2015}.
The population distribution of density profiles is governed by hyperparameters on each of the latent (or observed) parameters in $\bm{\theta}$ that themselves have weakly informative priors.
The prior distributions for the hyperparameters are given in \autoref{tab:abg_hyperpars}, and a graphical representation of the Bayesian model in the form of a directed acyclic graph is given in \autoref{fig:dag}.
To each latent parameter in $\bm{\theta}$ we assign a normal distribution, with mean $\mu_X$ and standard deviation $\sigma_X$ for the respective variable $X \in \bm{\theta}$.
To capture any covariance between latent parameters, we directly model the correlation matrix $\Omega= L_\mathrm{c}L_\mathrm{c}^\top$, where $L_\mathrm{c}$ is the Cholesky factor\footnote{The Cholesky factor of a positive-definite matrix is a lower-triangular matrix $L$ such that $\Sigma = LL^\top$.} of the correlation matrix.
We assign a \citet{lewandowski2009} prior on $L_\mathrm{c}$ as $L_\mathrm{c} \sim \mathrm{LKJCholesky}(\eta)$, with the prior on the correlation matrix thus becoming $p(\Omega | \eta) \propto \det(\Omega)^{\eta-1}$.
The variable $\eta$ controls the concentration of the prior about the identity matrix:~by setting $\eta=2$ (\autoref{tab:abg_hyperpars}), we mildly favor weak correlations while allowing moderate correlation structure. 

We define two new vectors of parameters:
\begin{align}
    \bm{\mu} &= \left[ \mu_{\log_{10}\rho_\mathrm{S}}, \mu_{\log_{10}r_\mathrm{S}}, \mu_{\log_{10}\alpha}, \mu_\beta, \mu_\gamma \right] \\
    \bm{L} &= \mathrm{diag}\left[ \sigma_{\log_{10}\rho_\mathrm{S}}, \sigma_{\log_{10}r_\mathrm{S}}, \sigma_{\log_{10}\alpha}, \sigma_\beta, \sigma_\gamma \right] \cdot L_\mathrm{c},
\end{align}
from which the latent parameter vector $\bm{\theta}$ is determined as
\begin{equation}
    \bm{\theta} = \bm{1} \cdot \bm{\mu}^\top + \bm{L}z.
\end{equation}

After the posterior sampling, we perform standard convergence and posterior-predictive checks to ensure the fits are robust by enforcing a diagnostic value $\hat{R}$ (comparison of between-chain and within-chain variance) of less than 1.05, a high effective sample, and that the number of divergent transitions is less than 1\% \citep[for details, see][]{vehtari2021,standevelopmentteam2018}.
We perform an additional prior sensitivity analysis using the power-scaling method described in \citet{kallioinen2024}, ensuring that the sensitivity diagnostic is less than 0.05 for the latent parameter $\bm{\theta}$ of the model. 
In doing so, we quantitatively ensure that the obtained posterior distributions are not strongly dependent on the assumed prior distributions.

As our model takes a number of free parameters, degeneracies between parameters are possible, thus highlighting the importance of using a Bayesian method to propagate parameter uncertainty. 
Typically, we find that the inner slope $\gamma$ is strongly negatively correlated with the characteristic density $\log_{10}\rho_\mathrm{S}$, as well as being correlated to a lesser degree with both $\alpha$ (positively) and $\beta$ (negatively).
Consequently, $\log_{10}\rho_\mathrm{S}$ tends to be negatively correlated with $\alpha$ and positively correlated with $\beta$.


\bibliography{ref}{}
\bibliographystyle{aasjournalv7}



\end{document}